\documentclass[12pt]{article}
\usepackage[dvips]{graphicx}
\usepackage{pdproc}

  \makeatletter 
  \def\@cite#1{[#1]} 
  \makeatother    
\begin{document}

\renewcommand{\thefootnote}{\alph{footnote}}

\title{
The Effect of Extra Dimension on Dark Matter \footnote{Presented by O.~Seto.}
}

\author{ NOBUCHIKA OKADA
}

\address{
Theory Division, KEK, Oho 1-1, Tsukuba, Ibaraki 305-0801, Japan
\\ {\rm E-mail: okadan@post.kek.jp}
}

\author{ OSAMU SETO
}

\address{
 Institute of Physics, National Chiao Tung University, \\
 Hsinchu, Taiwan 300, Republic of China
\\ {\rm E-mail: osamu@mail.nctu.edu.tw}
}

\abstract{
We investigate the thermal relic density of a cold dark matter
 in the brane world cosmology.
Since the expansion law in a high energy regime is modified
 from the one in the standard cosmology,
 if the dark matter decouples in such a high energy regime  
 its relic number density is affected by this modified expansion law.
We derive analytic formulas for the number density
 of the dark matter.
It is found that the resultant relic density is characterized
 by the ``transition temperature'' at which
 the modified expansion law in the brane world cosmology
 is connecting with the standard one,
 and can be considerably enhanced
 compared to that in the standard cosmology,
 if the transition temperature is low enough.
The implication to the neutralino dark matter also is mentioned.
}

\normalsize\baselineskip=15pt

\section{Introduction}

Recent various cosmological observations
 have established the $\Lambda$CDM cosmological model
 with a great accuracy,
 where the energy density in the present universe
 consists of about $73\%$ of the cosmological constant (dark energy),
 $23\%$ of non-baryonic cold dark matter and just $4\%$ of baryons.
However, to clarify the identity of the dark matter particle is still
 a prime open problem in cosmology and particle physics.

In the case that the dark matter is the thermal relic,
 we can estimate its number density by solving
 the Boltzmann equation
\begin{equation}
\frac{d n}{d t}+3Hn = -\langle\sigma v\rangle(n^2-n_{EQ}^2),
\label{n;Boltzmann}
\end{equation}
with the Friedmann equation,
\begin{equation}
H^2 = \frac{8\pi G}{3}\rho,
\label{FriedmannEq}
\end{equation}
where $H$ is the Hubble parameter,
 $n$ is the actual number density of dark matter particles,
 $n_{EQ}$ is the number density in thermal equilibrium,
 $\langle\sigma v\rangle$ is the thermal averaged product
 of the annihilation cross section $\sigma$ and the relative velocity $v$,
 $\rho$ is the energy density,
 and $G$ is the Newton's gravitational constant.
Eq. (\ref{n;Boltzmann}) is rewritten as
\begin{eqnarray}
\frac{d Y}{d x}
= -\frac{s\langle\sigma v\rangle}{xH}(Y^2-Y_{EQ}^2) ,
\label{Y;Boltzmann}
\end{eqnarray}
in terms of the number density to entropy ratio $Y = n/s$ and $x = m/T$.
As is well known, the final number density of dark matter particles 
 to entropy ratio is given as
\begin{equation}
Y(\infty) \simeq \frac{x_d}
{\lambda \left(\sigma_0 +\frac{1}{2} \sigma_1 x_d^{-1} \right)},
\label{Ystandard}
\end{equation}
with a constant 
\begin{equation}\lambda = \frac{xs}{H} 
 = 0.26 \left(\frac{g_{*S}}{g_*^{1/2}}\right) M_{P} m
\end{equation}
for models in which $\langle\sigma v\rangle$ is approximately
 parametrized as
 $\langle \sigma v\rangle = \sigma_0  + \sigma_1 x^{-1}
 +\mathcal{O}(x^{-2})$,
 where $g_*$ is the effective total number of relativistic degrees of freedom, 
 $x_d = m/T_d$, $T_d$ is the decoupling temperature
 and $m$ is the mass of the dark matter particle,  
 and $M_P \simeq 1.2 \times 10^{19}$GeV is the Planck mass \cite{Kolb}.

Recently, the brane world models have been attracting
 a lot of attention as a novel higher dimensional theory.
In these models, it is assumed that the standard model particles
 are confined on a ``3-brane'' while gravity resides in
 the whole higher dimensional spacetime.
The model first proposed by Randall and Sundrum (RS) \cite{RS},
 the so-called RS II model, is a simple and interesting one,
 and its cosmological evolution have been
 intensively investigated \cite{braneworld}.
In the model, our 4-dimensional universe is realized
 on the 3-brane with a positive tension
 located at the ultra-violet boundary
 of a five dimensional Anti de-Sitter spacetime.
The Friedmann equation for a spatially flat spacetime
 in the RS brane cosmology is found to be
\begin{equation}
H^2 = \frac{8\pi G}{3}\rho\left(1+\frac{\rho}{\rho_0}\right) ,
\label{BraneFriedmannEq}
\end{equation}
where $\rho_0 = 96 \pi G M_5^6$ 
 with $M_5$ being the five dimensional Planck mass, and we have omitted 
 the four dimensional cosmological constant and the so-called dark radiation.
The second term proportional to $\rho^2$ 
 is a new ingredient in the brane world cosmology
 and lead to a non-standard expansion law.

Here, we investigate the brane cosmological effect
 for the relic density of the dark matter
 due to this non-standard expansion law.
If the new terms in Eq.~(\ref{BraneFriedmannEq}) dominates
 over the term in the standard cosmology
 at the freeze out time of the dark matter,
 they can cause a considerable modification
 for the relic abundance of the dark matter \cite{OS}.

\section{Relic density in brane cosmology}

We are interested in the early stage in the brane world cosmology
 where the $\rho^2$ term dominates, namely, $\rho^2/\rho_0 \gg \rho$.
In this period, the coupling factor of collision term
 in the Boltzmann equation is given by
\begin{eqnarray}
\frac{s \langle\sigma v\rangle}{xH}
\simeq \frac{\lambda}{x_t^2} \langle\sigma v\rangle
 \label{collision-term}
\end{eqnarray}
where a new temperature independent parameter $x_t$ is defined as
\begin{equation}
 x_t^4 \equiv \left. \frac{\rho}{\rho_0}\right|_{T=m}.
\end{equation}
Note that the evolution of the universe can be divided into two eras.
At the era $x \ll x_t$ the $\rho^2$ term
 in Eq.~(\ref{BraneFriedmannEq}) dominates
 (brane world cosmology era),
 while at the era $x \gg x_t$ the expansion law
 obeys the standard cosmological law (standard cosmology era).
In the following, we call the temperature defined
 as $T_t = m x_t^{-1}$ (or $x_t$ itself) ``transition temperature''
 at which the evolution of the universe changes
 from the brane world cosmology era to the standard cosmology era.
We consider the case that
 the decoupling temperature of the dark matter particle
 is higher than the transition temperature.

At the early time, the dark matter particle is
 in the thermal equilibrium and $Y=Y_{EQ}+\Delta$ tracks $Y_{EQ}$ closely.
After the temperature decreases, the decoupling occurs at $x_d$
 roughly evaluated as $\Delta(x_d) \simeq Y(x_d) \simeq Y_{EQ}(x_d)$.
The solutions of the Boltzmann equation during the $x_t>x>x_d$ epoch are given as
\begin{eqnarray}  
  && \frac{\lambda \sigma_0}{x_t^2} \left( x - x_d \right),
   \nonumber \\
\frac{1}{\Delta(x)}-\frac{1}{\Delta(x_d)} =
 &&  \frac{\lambda \sigma_1}{x_t^2} \ln\left( \frac{x}{x_d} \right),
  \nonumber \\
 &&  \frac{\lambda \sigma_n}{x_t^2}
   \left( \frac{1}{n-1}\right)
\left( \frac{1}{x_d^{n-1}} - \frac{1}{x^{n-1}}
\right),
 \label{solutions}
\end{eqnarray}  
 for $n=0$ (S-wave process), $n=1$ (P-wave process), and $n >1$ respectively.
Here we have parametrized $\langle \sigma v \rangle$ 
 as $\langle \sigma v \rangle = \sigma_n x^{-n}$.
Note that $\Delta(x)^{-1}$ is continuously growing 
 without saturation for $n \leq 1$. 
This is a very characteristic behavior 
 of the brane world cosmology, 
 comparing the case in the standard cosmology 
 where $\Delta(x)$ saturates after decoupling 
 and the resultant relic density is roughly 
 given by $Y(\infty) \simeq Y(x_d)$. 
For a large $x \gg x_d$ in Eq.~(\ref{solutions}), 
 $\Delta(x_d)$ and $x_d$ can be neglected. 
When $x$ becomes large further and reaches $x_t$, 
 the expansion law changes into the standard one, 
 and then $Y$ obeys the Boltzmann equation 
 with the standard expansion law for $x \geq x_t$. 
Since the transition temperature is smaller 
 than the decoupling temperature in the standard cosmology 
 (which case we are interested in), 
 we can expect that the number density freezes out 
 as soon as the expansion law changes into the standard one. 
Therefore the resultant relic density can be roughly 
 evaluated as $Y(\infty) \simeq \Delta(x_t)$ 
 in Eq.~(\ref{solutions}). 
 
Now, we show analytic formulas of the final relic density
 of the dark matter in the brane world cosmology.
For $n=0$, we find the resultant relic density
\begin{equation}
 Y(\infty) \simeq 0.54 \frac{x_t}{\lambda \sigma_0}.
\end{equation}
 in the case of $x_d \ll x_t$.
Note that the density is characterized
 by the transition temperature $x_t$ as we expected.
By using the well known formula (\ref{Ystandard}),
 for a given $\langle\sigma v\rangle$,
 we obtain the ratio of the energy density of the dark matter
 in the brane world cosmology ($\Omega_{(b)}$)
 to the one in the standard cosmology ($\Omega_{(s)}$)
 such that
\begin{eqnarray}
\frac{\Omega_{(b)}}{\Omega_{(s)}}
\simeq 0.54 \left(\frac{x_t}{x_{d(s)}}\right) ,
\end{eqnarray}
where $x_{d(s)}$ is the decoupling temperature
 in the standard cosmology.
Similarly, for $n=1$ we find the resultant relic density
\begin{eqnarray}
Y(\infty) \simeq \frac{x_t^2}{\lambda \sigma_0 \ln x_t}
\end{eqnarray}
in the case of $x_d \ll x_t$.
Then, the ratio of the energy density of the dark matter
 is found to be
\begin{eqnarray}
\frac{\Omega_{(b)}}{\Omega_{(s)}}
\simeq \frac{1}{2 \ln x_t}
 \left(\frac{x_t}{x_{d(s)}}\right)^2 .
\end{eqnarray}
We can obtain results for the case of $n >1$
 in the same manner.  

\section{Summary}

We have investigated the thermal relic density
 of the cold dark matter in the brane world cosmology.
If the five dimensional Planck mass is small enough,
 the $\rho^2$ term in the modified Friedmann equation
 can be effective when the dark matter is decoupling.
We have derived the analytic formulas for the relic density
 and found that the resultant relic density can be enhanced.
The enhancement factor is characterized
 by the transition temperature, at which
 the evolution of the universe changes from the brane world cosmology era
 to the standard cosmology era.
 
If our scenario is applied to the supersymmetric dark matter 
 such as the neutralino, its aboundance could be enhanced. 
 Thus a larger annihilation cross section, as in the 
 higgsino-like case, may be favorable 
 if the $\rho^2$ term was significant at the decoupling time.
Allowed regions obtained in the previous analysis
 in the standard cosmology \cite{CDM}
 would be dramatically modified in such cases.

\section{Acknowledgements}

We would like to thank the organizers of this conference for giving us an 
opportunity to present this work.
N.O. is supported in part by the Grant-in-Aid for 
Scientific Research  (\#15740164). 
O.S. is supported by the National Science Council of Taiwan 
under the grant No. NSC 92-2811-M-009-021. 

\bibliographystyle{plain}

\end{document}